\documentclass[a4paper, aps, tightenlines, nofootinbib, 12pt]{revtex4}
\usepackage{hyperref}
\usepackage{amsmath}
 \usepackage{bbm}
\newsavebox{\myhbar}
\savebox{\myhbar}{$\hbar$}
\usepackage{graphicx}
\usepackage{enumerate}
\usepackage{mdframed}
\usepackage{braket}
\usepackage{url}
\usepackage{wasysym}

\usepackage{subfig}
\usepackage{graphicx}
\graphicspath{ {./images/} }
\usepackage{float}
\renewcommand*{\hbar}{\mathalpha{\usebox{\myhbar}}}
 \begin{document}

 \title{Relativistic Quantum Thermodynamics of Moving  Systems} \author {  Nikolaos Papadatos}
 \email{n.papadatos@upatras.gr}
  \author{ Charis Anastopoulos}
  \email{anastop@upatras.gr}
 \affiliation{ Department of Physics, University of Patras, 26500 Patras,  Greece}

\begin{abstract}
We analyse the   thermodynamics of a  quantum system in a trajectory of constant velocity that interacts with a static thermal bath. The latter is modeled by a massless scalar field in a thermal state. We consider  two different couplings of the moving system to the heat bath, a coupling of the Unruh-DeWitt type and a coupling that involves the time derivative of the field. We derive the master equation for the reduced dynamics of the moving quantum system. It  has the  same form with the quantum optical master equation, but with different coefficients that depend on velocity. This master equation  has a unique asymptotic state for each type of coupling, and it is characterized by a well-defined notion of heat-flow. Our analysis of the second law of thermodynamics  leads to a surprising equivalence: a moving heat bath is physically equivalent to a mixture of  heat baths at rest, each with a different temperature. There is no unique rule for the Lorentz transformation of temperature. We propose that Lorentz transformations of thermodynamic states are well defined in an extended thermodynamic space that is obtained as a  convex hull of the standard thermodynamic space.
\end{abstract}

\maketitle

\pagebreak

\section{Introduction}

\subsection{Motivation}
In this paper, we analyse the thermodynamics of a quantum system  in a trajectory of constant velocity that  interact with a static thermal bath.  This system turns out to have a well-defined notion of heat flow. More importantly, it manifests an intriguing property. The heat bath, in motion in the reference frame of the quantum system, is equivalent to
a continuum of  heat baths at rest with respect to the moving system, each with a different temperature.

This paper is partly motivated by the long-standing puzzle of the relativistic transformations of temperature. This puzzle  originates from the early days of special relativity. Von Mosengeil \cite{vMos},  Planck \cite{Planck} and Einstein \cite{Einstein}   analyzed Lorentz transformations for  a body of temperature $T$ in its rest frame. They proposed that the temperature $T'$ for an observer that moves at velocity $v$ with respect to the  body is  $T' = T\sqrt{1-v^2} < T$. This view was accepted for more than 50 years, until it was challenged by Ott \cite{Ott}.  He argued that $T' = T/\sqrt{1-v^2} > T$. Arzeli\'es agreed with Ott's formula, but proposed a different law of transformation for the internal energy \cite{Arzel}.
A few years later, Landsberg proposed that the temperature is a Lorentz scalar, i.e.,  $T' = T$ \cite{Landsberg}.  Cavalleri and Salgarelli agreed that the temperature is a scalar, but they asserted that the Clausius formula $dQ = TdS$ holds only in a system's rest frame \cite{CaSa}. In their perspective, the familiar rules of thermodynamics are valid only in the rest frame.

The debate about the correct transformation rules for thermodynamic variables has been raging ever since the 1960s, and still there is no consensus---for reviews see, Ref. \cite{reviews}. Part of the problem is the lack of concrete operational ways of measuring thermodynamic quantities in moving systems. In 1995, Costa and Matsas \cite{CoMa} proposed a non-thermodynamic type of measurement: an Unruh-DeWitt (UdW) detector moving with velocity $v$ and interacting with a thermal scalar field. An UdW detector \cite{Unruh76, Dewitt, HLL12} is a quantum system that moves along a classical trajectory, and interacts with a quantum field through a dipole coupling. The UdW model and its generalizations originate from studies of particle creation in black holes and in accelerated reference frames.

The expression derived in Ref.  \cite{CoMa} for the detection rate of an UdW detector can not be brought into the characteristic Planckian form of black-body radiation.  Landsberg and Matsas took this as an indication that a universal and continuous Lorentz transformation of temperature is impossible \cite{LaMa}---see, also \cite{Naka} for a critique.

However, a particle detector is not a thermometer. The expression for the detection rate that was employed in Ref. \cite{CoMa}  follows from time-dependent perturbation theory, and it is valid at times much smaller than the relaxation time of the detector. It describes a system that has not come into equilibrium with the field bath. In contrast, ordinary thermometers give reliable records only after they have established equilibrium with  the larger system.

The ambiguous use of the word `detector' is often a source of confusion. If a detector is literally viewed as a measuring apparatus that records particles---and leaves a macroscopic measurement record---
 then the perturbative expression for the detection rate  suffices \cite{AnSav11}. However, the UdW Hamiltonian can  also be used to describe a small microscopic system  that interacts with a quantum field, for example, an atom or a molecule. Rather misleadingly, these systems are also referred to as detectors, even if they  leave no measurement records.

 The latter type of systems must be described by the theory of open quantum systems \cite{Davies, BrePe07}, which takes into account dissipation, noise and backreaction effects. In open quantum systems, time evolution is described by a master equation, which enables a thermodynamic analysis \cite{Kosloff, Alicki}. For accelerated UdW detectors  and the implications to the Unruh effect in this context, see, \cite{Benatti, MouAn, Mou2}.

We will use the theory of open quantum systems in order  to analyze the dynamics of moving thermometers, and more generally, the interaction of  moving quantum thermodynamic systems with thermal reservoirs.

\subsection{Analysis and results}
Our model consists of a small quantum  system (denoted by S). We will refer to S as a quantum probe. The
 center of momentum (CoM) of S moves at a constant velocity with respect to a reference frame $\Sigma$.
The Hamiltonian of S  internal degrees of freedom i.e., degrees of freedom that are not related to motion of the CoM.  The probe interacts with a massless scalar field $\hat{\phi}(x)$ at a thermal state of temperature $T$ with respect to $\Sigma$.

We consider two different couplings of the small system to the field bath. First, the standard UdW coupling, where $S$ couples to the field operator $\hat{\phi}(x)$, and second, a coupling of $S$ to the time derivative of  $\hat{\phi}(x)$. The latter coupling provides a more accurate representation of the dipole interaction  of atoms to the electromagnetic field.

Then, we construct the second-order master equation that describes the reduced dynamics of the system S. Past analyses of the open system dynamics of moving UdW detectors have shown that non-Markovian effects may be important \cite{HuLin, MouAn, Mou2, SLMV}, but  their contribution is small at long times. Hence, the Markovian approximation inherent in the second-order master equation suffices for describing the approach to equilibrium.

Our results are the following.
\begin{enumerate}[(i)]
\item The system S evolves according to the quantum-optical master equation. The only difference is that the coefficients of the master equation that represent to the mean number of bath quanta   do not have the usual Planckian form.
\item There exists unique asymptotic states for many choices of self-dynamics for S. The asymptotic states  do not depend on the initial state of the system or on the strength of the system-field interaction (i.e., the dissipation rate). Still, they are not universal: each type of coupling leads to a different asymptotic state.
\item  The average energy in the rest frame of S  defines an empirical temperature for the bath.
\item There is a well-defined notion of "hotness", i.e., we can say when the moving probe is hotter or colder than the bath,  as reflected in the direction of the heat flow.
\item The second law of thermodynamics is satisfied, i.e., entropy production is positive.

\item The moving heat bath is equivalent to a continuum of heat baths in the rest frame of S, with temperatures $T'$ in the range
\begin{eqnarray}
T \frac{1-|v|}{1+|v|} \leq T' \leq T \frac{1+|v|}{1-|v|}.
\end{eqnarray}
The contribution of each heat bath at temperature $T'$  to entropy production is weighted by a probability distribution that depends on the type of coupling.
\end{enumerate}

\medskip

Evidently, there is no Lorentz transformation for temperature. A static observer cannot assign a unique absolute temperature to a moving heat bath.   Nonetheless, the moving heat bath still behaves like a thermodynamic system, albeit with significant differences from ordinary thermodynamics. First, we cannot assign a unique absolute temperature to the bath, even if the empirical temperatures are well defined. Second, asymptotic states for a given temperature and velocity are not universal.

A spacetime-covariant thermodynamic description likely requires additional physical observables, in order to define an extended thermodynamic state space.
We propose that the extended state space contains  convex combination of ordinary thermodynamic states, for example,  convex combinations $\sum_i c_i \hat{\rho}_{\beta_i}$ of Gibbs states $\hat{\rho}_{\beta_i}$ at different temperatures $\beta_i$. This idea appears natural in the context of our results. However, further analysis is required, in order to test this proposal and in order to formulate a consistent relativistic thermodynamics for quantum systems.

\bigskip

The structure of this paper is the following. In Sec. 2, we derive the master equation for a  moving quantum system that interacts with  a massless scalar field in a thermal state. In Sec. 3, we identify the asymptotic states for simple choices of the Hamiltonian. In Sec. 4, we undertake a thermodynamic analysis of the master equation with an  emphasis on the three laws of thermodynamics. In Sec. 4, we propose a rule for the relativistic transformation of thermodynamic states, and we construct the associated state space. In the final section, we discuss future directions.

\section{Master equation for a moving quantum system that interacts with  a thermal bath}

\subsection{The model}
 We consider a composite quantum system that consists of a microscopic probe S and a quantum scalar field. The system  is described by a Hilbert space ${\cal H}_S \otimes {\cal H}_{\phi}$, where ${\cal H}_S $ is the Hilbert space of the probe and ${\cal H}_{\phi}$ is the Hilbert space of the field.

The Hamiltonian is
\begin{eqnarray}
\hat{H}=\hat{h}\otimes\hat{I}+\hat{I}\otimes\hat{H}_{\phi}+\hat{V} \label{Ham1}
\end{eqnarray}
where $\hat{h}$ is the Hamiltonian that generates time translations with respect to the proper time parameter $\tau $ of S.


The Hamiltonian $\hat{H}_{\phi}$ describes a free massless scalar field,
\begin{equation}
 \hat{H}_\phi= \frac{1}{2} \int d^3x\Big(\hat{\pi}^2+(\nabla\hat{\phi})^2\Big)
\end{equation}
where $\hat{\pi}({\pmb x})$ is the conjugate momentum of the field $\hat{\phi}({\pmb x})$. The Heisenberg picture fields are defined as $\hat{\phi}(X):= e^{i\hat{H}_{\phi}t} \hat{\phi}({\pmb x})  e^{-i\hat{H}_{\phi}t} $, where $x = (t, {\pmb x})$.

The interaction term is  of the general form,
\begin{equation}
\hat{V} = \lambda  \hat{A} \otimes \hat{O}({\pmb x}(\tau)), \label{coupling}
\end{equation}
where $\lambda$ is a coupling constant, $\hat{A}$ is a self-adjoint operator on ${\cal H}_S$ and ${\pmb x}(\tau)$ is the  path of the detector; $\hat{O}(x)$ is a local composite operator for the scalar field.
In the interaction picture,  $\hat{V} $  involves the field operator $\hat{O}({\pmb x}(\tau))$. In this paper, we   focus on trajectories
\begin{eqnarray}
\label{trajectories}
x(\tau) = (\cosh u, \sinh u , 0, 0) \;  \tau,
\end{eqnarray}
where $u$ is the rapidity of the trajectory, with associated velocity $v = \tanh u$.

We assume a factorized  initial state $\hat{\rho}_0 \otimes \hat{\rho}_{\phi}$. We consider a
 stationary, space-translation-invariant, and rotation-invariant state for the quantum field, eventually to be identified with a Gibbs state at temperature $\beta^{-1}$,
 \begin{eqnarray}
 \hat{\rho}_{\phi} = \frac{e^{- \beta \hat{H}_{\phi}}}{Tr e^{- \beta \hat{H}_{\phi}}}
 \end{eqnarray}

  The master equation for the probe depends on the Wightman function \cite{Wel00}
\begin{eqnarray}
G(x) := Tr \left[ \hat{O}(x)\hat{O}(0) \hat{\rho}_{\phi} \right]. \label{Wightman0}
\end{eqnarray}

In this paper, we will consider rotation-invariant coupling operators $\hat{O}(x)$. In particular, we will analyze two cases:
\begin{itemize}
\item $\hat{O}(x) = \hat{\phi}(x)$. This is the usual Unruh-deWitt (UdW) coupling;
\item $\hat{O}(x) = \dot{\hat{\phi}}(x)$. This time-derivative (TD)  coupling  best simulates the electromagnetic dipole interaction\footnote{The electromagnetic dipole coupling is of the form ${\pmb d} \cdot {\pmb E}$, where ${\pmb d}$ is the dipole moment, and ${\pmb E} = \dot{{\pmb A}}$ is the electric field. If we assume that the direction of the dipole moment fluctuates homogeneously, then the Wightman function for the EM field coincides with that of the scalar field with TD coupling, modulo a multiplicative factor.}.
\end{itemize}

The Wightman function (\ref{Wightman0}) for the UdW coupling takes the form
 \begin{eqnarray}
 G_{UdW}(x) = G_0(x) + \frac{1}{4 \pi^2r} \int_0^{\infty} dk n_{k}\left[\sin[k(t+r)] - \sin[k(t-r)]\right], \label{wightman}
 \end{eqnarray}
 where $r = |{\pmb x}|$,
 \begin{eqnarray}
 G_0(x) = - \lim_{\epsilon \rightarrow 0^+} \frac{1}{4\pi^2\left[ (t- i \epsilon)^2-r^2\right]},
 \end{eqnarray}
is the Wightman function of the vacuum, and $n_k$ is the expected number of particles of momentum ${\pmb k}$. Due to spherical symmetry $n_k$ depends only on $k = |{\pmb k}|$. For a Gibbsian field state, $n_k = (e^{\beta k}-1)^{-1}$.

The Wightman function (\ref{Wightman0}) for the TD coupling is
\begin{eqnarray}
G_{TD}(x) = - \frac{\partial^2}{\partial t^2}  G_{UdW}(x).
\end{eqnarray}

In the open quantum systems description, the effect of the environment is contained in the bath two-time correlation function. In the present context, the bath correlation function coincides with the Wightman function $G[x(\tau) - x(\tau')]$ evaluated at a pair of points $x(\tau)$ and $x(\tau')$ along the trajectory of the probe. For paths given by Eq. (\ref{trajectories}), the bath correlation function is static, i.e., $G[x(\tau) - x(\tau')]$ is a function $g(\tau - \tau')$ only of the difference $\tau - \tau'$ \cite{Letaw}.

For the UdW coupling,

\begin{eqnarray}
g_{UdW}(\tau) = - \lim_{\epsilon \rightarrow 0^+} \frac{1}{4\pi^2 (\tau- i \epsilon)^2 } + \frac{1}{4 \pi^2 |\tau| \sinh u  } \int_0^{\infty} dk n_{k}\left[\sin(e^{u} k \tau) - \sin(e^{-u}k \tau)\right]. \label{gt1}
\end{eqnarray}
For the TD coupling,
\begin{eqnarray}
g_{TD}(\tau) =  \lim_{\epsilon \rightarrow 0^+} \frac{1+ 2 \cosh(2u)}{2\pi^2 (\tau- i \epsilon)^4 }+ \frac{1}{4 \pi^2 |\tau| \sinh u  } \int_0^{\infty} dk k^2 n_{k}\left[\sin(e^{u} k \tau) - \sin(e^{-u}k \tau)\right]. \label{gt2}
\end{eqnarray}

\subsection{The master equation}
Given the Hamiltonian (\ref{Ham1}) and the factorized initial states, the reduced dynamics of the probe is expressed in terms of the {\em second order master equation}. The master equation can be derived in different ways. One way involves the van Hove limit, i.e.,   the limit $\lambda \rightarrow 0$  with $\lambda^2 t$  fixed \cite{Davies}. A different approach involves the successive use of the Born, Markov and Rotating Wave Approximations (RWA) \cite{BrePe07}.

To proceed, we  define the {\em transition operators}

\begin{eqnarray}
\hat{A}_{\omega} = \sum_{n, m, \epsilon_m - \epsilon_n = \omega}  \langle n|\hat{A}|m\rangle  |n\rangle\langle m|,
\end{eqnarray}
indexed by the set of all possible energy differences $\omega =  \epsilon_m - \epsilon_n$.

By construction,   transition operators satisfy the identities
\begin{eqnarray}
\sum_\omega \hat{A}_{\omega} = \hat{A}, \hspace{1cm}
\hat{A}_{-\omega} = \hat{A}_{\omega}^{\dagger},
\end{eqnarray}
and the commutation relations
\begin{eqnarray}
\left[\hat{h}, \hat{A}_{\omega}\right] &=& - \omega  \hat{A}_{\omega}, \\
\left[\hat{h}, \hat{A}^{\dagger}_{\omega}\right] &=&  \omega  \hat{A}_{\omega},\\
\left[\hat{h}, \hat{A}^{\dagger}_{\omega}\hat{A}_{\omega}\right] &=& \left[\hat{h}, \hat{A}_{\omega}\hat{A}^{\dagger}_{\omega}\right] =  0.
\end{eqnarray}

The second-order master equation for the reduced density matrix $\hat{\rho}$ of the probe is
\begin{eqnarray}
\frac{\partial \hat{\rho}}{\partial \tau} = - i [ \hat{h}, \hat{\rho}] +   \lambda^2 \sum_{\omega} \tilde{g}(\omega)  \left( \hat{A}_{\omega}\hat{\rho}\hat{A}^{\dagger}_{\omega}
- \hat{A}^{\dagger}_{\omega} \hat{A}_{\omega}\hat{\rho} \right) \nonumber +\\
+\lambda^2 \sum_{\omega} \tilde{g}^*(\omega) \left(  \hat{A}^{\dagger}_{\omega}\hat{\rho}\hat{A}_{\omega}  - \hat{\rho} \hat{A}^{\dagger}_{\omega} \hat{A}_{\omega}\right),
\end{eqnarray}
where
\begin{eqnarray}
\tilde{g}(\omega) = \int_0^{\infty}  d \tau e^{i \omega \tau} g(\tau).
\end{eqnarray}

We split $\tilde{g}(\omega)$ into its real and imaginary part while  absorbing the constant $\lambda^2$: $\lambda^2 \tilde{g}(\omega) = \frac{1}{2} \Gamma(\omega) + i \Delta(\omega)$, where
\begin{eqnarray}
\Gamma(\omega) = \lambda^2 [\tilde{g}(\omega) + \tilde{g}^*(\omega)] = 2 \lambda^2  \int_{0}^{\infty} d \tau \cos (\omega \tau) g(\tau), \\
\Delta(\omega) = \frac{\lambda^2 }{2i} \left[ \tilde{g}(\omega) - \tilde{g}^*(\omega) \right] = \lambda^2 \int_{0}^{\infty} d \tau \sin (\omega \tau) g(\tau) .
\end{eqnarray}

Then, the master equation becomes

\begin{eqnarray}
\frac{\partial \hat{\rho}}{\partial \tau} = - i [ \hat{h} + \hat{h}_{LS}, \hat{\rho}] +    \sum_{\omega} \Gamma(\omega)  \left[ \hat{A}_{\omega}\hat{\rho}\hat{A}^{\dagger}_{\omega} -\frac{1}{2}  \hat{A}^{\dagger}_{\omega}\hat{A}_{\omega}\hat{\rho}  -  \frac{1}{2}  \hat{\rho} \hat{A}^{\dagger}_{\omega} \hat{A}_{\omega}\right],
\end{eqnarray}
where
\begin{eqnarray}
\hat{h}_{LS} :=  \sum_{\omega} \Delta(\omega)  \hat{A}^{\dagger}_{\omega} \hat{A}_{\omega},
\end{eqnarray}
is a correction to the Hamiltonian, that implements a Lamb-shift of the energy levels.

Eqs. (\ref{gt1}) and (\ref{gt2}) imply that $\Gamma(\omega)$ is of the form
\begin{eqnarray}
\Gamma(\omega) = \gamma(|\omega|) \left\{ \begin{array}{cc} 1+ N(\omega),& \omega > 0\\N(|\omega|),&\omega< 0\end{array}\right.,
\end{eqnarray}
where the explicit form of $\gamma(\omega)$ and $N(\omega)$   will be given in Sec. 2.C.

The master equation   becomes
\begin{eqnarray}
\frac{\partial \hat{\rho}}{\partial \tau} = - i [ \hat{h} + \hat{h}_{LS}, \hat{\rho}] +    \sum_{\omega > 0 } \gamma(\omega) [N(\omega)+1]  \left[ \hat{A}_{\omega}\hat{\rho}\hat{A}^{\dagger}_{\omega} -\frac{1}{2}  \hat{A}^{\dagger}_{\omega}\hat{A}_{\omega}\hat{\rho}  -  \frac{1}{2}  \hat{\rho} \hat{A}^{\dagger}_{\omega} \hat{A}_{\omega}\right]\nonumber \\
+     \sum_{\omega > 0 } \gamma(\omega) N(\omega)  \left[ \hat{A}^{\dagger}_{\omega}\hat{\rho}\hat{A}_{\omega} -\frac{1}{2}  \hat{A}_{\omega}\hat{A}^{\dagger}_{\omega}\hat{\rho}  -  \frac{1}{2}  \hat{\rho} \hat{A}_{\omega} \hat{A}^{\dagger}_{\omega}\right]. \label{mastereq}
\end{eqnarray}

 Eq. (\ref{mastereq}) is of the same form with the quantum optical master equation \cite{BrePe07}. The only difference is that the expected  number of quanta $N(\omega)$ is not given by the Planck distribution.

\subsection{The coefficients in the master equation}

\noindent {\bf  UdW coupling:} The coefficients $\gamma(\omega)$ and $N(\omega)$ are defined for positive $\omega$ as
\begin{eqnarray}
\gamma_{UdW}(\omega) &=& \frac{\lambda^2}{2\pi} \omega \\
N_{UdW}(\omega) &=& \frac{1}{2 \omega \sinh u} \int_{\omega e^{-u}}^{\omega e^u} n_k dk. \label{Nomega}
\end{eqnarray}

In deriving Eq. (\ref{Nomega}), we used the identity
\begin{eqnarray}
\int_0^{\infty} d \tau \cos(\omega \tau) \sin(a\tau) \tau^{-1} = \frac{\pi}{4} \left[ \mbox{sgn}(a-\omega) + \mbox{sgn}(a+\omega)\right].
\end{eqnarray}

For an initial  thermal state of the field, Eq. (\ref{Nomega}) yields
\begin{eqnarray}
N_{UdW}(\omega) = \frac{1}{2\beta \omega \sinh u} \log \frac{ 1 - e^{-\beta \omega e^{u}}}{1 - e^{-\beta \omega e^{-u}}}. \label{Nomega2}
\end{eqnarray}

The function $N_{UdW}(\omega)$ has the following asymptotic  behavior.
\begin{enumerate}
\item Low-velocity regime, $|u| << 1$: $N_{UdW}(\omega) =  n_{\omega} + \frac{1}{2} \omega n'_{\omega} u^2 + \ldots $.
\item High-velocity regime, $|u| >> 1$: $N_{UdW}(\omega) = \frac{1}{\beta \omega} e^{-|u|} [|u| - \log(\beta \omega)]$.
\item Low-temperature regime, $\beta \omega e^{-|u|} >> 1$:  $N_{UdW}(\omega) = \frac{e^{-\beta \omega e^{-|u|}}}{2 \beta \omega \sinh |u|}$.
\item High-temperature regime, $\beta \omega e^{|u|} << 1$: $N_{UdW}(\omega) =  \frac{1}{\beta \omega} \frac{u}{\sinh u}$.
\end{enumerate}
In comparison, the Planck distribution $n_{\omega}$ behaves as $e^{-\beta \omega}$ for $\beta \omega >>1$ and as $\frac{1}{\beta \omega}$ for  $\beta \omega <<1$. It follows that
\begin{itemize}
\item $N_{UdW}(\omega) < n_{\omega}$, as $\beta \omega \rightarrow 0$, \\
\item  $N_{UdW}(\omega) > n_{\omega}$, as $\beta \omega \rightarrow \infty$.
\end{itemize}

\medskip

The contribution to the Lamb shift is
\begin{eqnarray}
\Delta_{UdW}(\omega) = \mbox{sgn}(\omega)\left[ \Delta_0 + \frac{\lambda^2}{8\pi^2 \sinh u} \int_0^{\infty} dk n_k  \log \left| \frac{(\omega + ke^{-u})(\omega-k e^{u})}{(\omega - ke^{-u})(\omega+k e^{u})}\right|\right],
\end{eqnarray}
where the integral $\Delta_0 =  -\frac{\lambda^2}{4\pi^2}\int_0^{\infty} d \tau \sin(|\omega| \tau)/\tau^2$ diverges at $\tau = 0$. We regularize by introducing a cut-off $\epsilon$ in the lower range of integration. Then,
\begin{eqnarray}
\Delta_0 =  \frac{\lambda^2 |\omega|}{4\pi^2} \log(e^{\gamma-1} |\omega| \epsilon),
\end{eqnarray}
 where here $\gamma$ stands for the Euler-Macheronni constant.

\bigskip

\noindent {\bf  TD coupling:}
 The coefficients $\gamma(\omega)$ and $N(\omega)$ are defined for positive $\omega$ as
\begin{eqnarray}
\gamma_{TD}(\omega) &=& \frac{\lambda^2[1+2 \cosh(2u)]}{6\pi} \omega^3 \\
N_{TD}(\omega) &=& \frac{3}{2 \omega^3 \sinh u [1+2 \cosh(2u)]} \int_{\omega e^{-u}}^{\omega e^u} n_k k^2 dk. \label{Nomega2b}
\end{eqnarray}
For an initial  thermal state of the field, Eq. (\ref{Nomega}) yields
\begin{eqnarray}
N_{TD} (\omega) = \frac{3}{2\beta^3 \omega^3 \sinh u [1+2 \cosh(2u)]} [F(\beta \omega e^{-u}) - F(\beta \omega e^{u})], \label{Nomega3}
\end{eqnarray}
where the function
\begin{eqnarray}
F(x) = 2g_3(e^{-x}) + 2 x g_2(e^{-x}) + x^2 g_1(e^{-x})
\end{eqnarray}
 is expressed in terms of the polylogarithm functions $g_{\ell}(x) = \sum_{n=1}\frac{x^n}{n^{\ell}}$. Note that $g_1(x) = - \log(1-x)$, and that $g_{\ell}(1) = \zeta(\ell)$, where $\zeta$ is Riemann's zeta function.

The function $N_{TD}(\omega)$ has the following asymptotic  behavior.
\begin{enumerate}
\item Low-velocity regime, $|u| << 1$: $N_{TD}(\omega)  =  n_{\omega} + \left(\frac{1}{2} \omega n'_{\omega}  + \frac{2}{3} n_{\omega}\right) u^2+ \ldots $.
\item High-velocity regime, $|u| >> 1$: $N_{TD}(\omega) =  \frac{6 \zeta(3)}{\beta^3 \omega^3} e^{- 3|u|}$.
\item High-frequency regime, $\beta \omega e^{-|u|} >> 1$:  $N_{TD}(\omega) = \frac{3e^{-\beta \omega e^{-|u|}}}{2 \beta \omega \sinh |u|[1+\cosh(2u)]} $.
\item Low-frequency regime, $\beta \omega e^{|u|} << 1$: $N_{TD}(\omega) =  \frac{1}{\beta \omega} \frac{3\cosh u}{1+2 \cosh(2u)} $.
\end{enumerate}

\medskip

For small $u$, $N_{TD}(\omega)$ is practically indistinguishable from $N_{UdW}(\omega)$. For large $u$, $N_{TD}(\omega)$ is significantly smaller than $N_{UdW}(\omega)$. This behavior is demonstrated graphically  in Fig.1.

The contribution to the Lamb shift is
\begin{eqnarray}
\Delta_{TD}(\omega) = \mbox{sgn}(\omega)\left[ \tilde{\Delta}_0 + \frac{\lambda^2}{8\pi^2 \sinh u} \int_0^{\infty} dk k^2 n_k  \log \left| \frac{(\omega + ke^{-u})(\omega-k e^{u})}{(\omega - ke^{-u})(\omega+k e^{u})}\right|\right],
\end{eqnarray}
where the integral
\begin{eqnarray}
\tilde{\Delta}_0 =  \frac{\lambda^2[1+2\cosh(2u)]}{2\pi^2}\int_0^{\infty} d \tau \sin(|\omega| \tau)/\tau^4
\end{eqnarray}
 diverges at $\tau = 0$. We regularize by introducing a cut-off $\epsilon$ in the lower range of integration. As $\epsilon \rightarrow 0$,
\begin{eqnarray}
\tilde{\Delta}_0 =\frac{\lambda^2 [1+2\cosh(2u)]|\omega|^3}{12\pi^2} \left[\frac{3}{(\omega \epsilon)^2} +\log(|\omega| \epsilon e^{\gamma-1}) \right].
\end{eqnarray}

\begin{figure}[H]
    \centering
   {{\includegraphics[width=11cm]{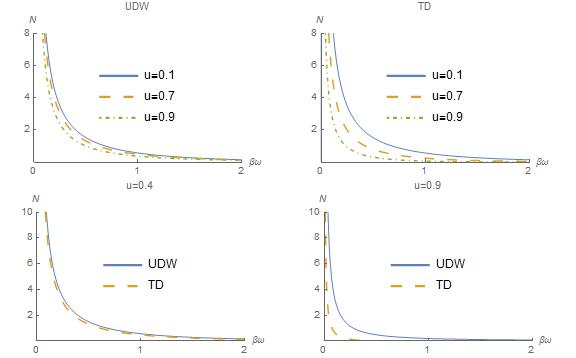} }}%
  \caption{ \small (i) $N_{UdW}$ as a function of $\beta \omega$ for different values of $u$. (ii)$N_{TD}$ as a function of $\beta \omega$ for different values of $u$. (iii)  $N_{UdW}$ vs $N_{TD}$ as a function of $\beta \omega$, for $u = 0.4$. (iv) $N_{UdW}$ vs $N_{TD}$ as a function of $\beta \omega$, for $u = 0.9$.}
\end{figure}

\subsection{Special limits}

{\em Non-relativistic limit.} This limit corresponds to the regime $|u|<< 1$. By the analysis of the previous section, $N(\omega) = n_{\omega} + O(|u|^2)$, and we recover the standard optical master equation for a heat bath at temperature $T$. The first correction is of order $O(|u|^2)$ and it depends on the type of coupling.

\smallskip

{\em Ultra-relativistic limit.} This limit corresponds to   $|u|\rightarrow \infty$. In this regime, $N(\omega)$ is suppressed exponentially $e^{n|u|}$ for some power $n$. Hence, very fast probes see the field as though it is in its ground state.

\small

\smallskip {\em Classical limit.} The function $N(\omega)$ depends on $\hbar$ only through the combination $\hbar \omega$. This means that there is no classical limit ($\hbar \rightarrow 0$). This is not surprising, the classical limit does not exist even for the usual quantum optical master equation. The reason is that there is no classical thermal state for the electromagnetic field: the classical statistical mechanics of the electromagnetic field is not consistent.

Hence, there is no way to relate our results to  derivations of the transformation laws for temperature that are based on classical physics. To this end, it is necessary to consider a thermal environment  with a good classical limit. This is the case, for example, of a bosonic or fermionic ideal gas, with particles of finite mass. This reduces to a classical ideal gas in the appropriate regime. The challenge there is to construct a physically meaningful coupling between the quantum probe and the environment. We will undertake this work in a future publication. Note that even in such systems, there is no {\em a priori} guarantee of a correspondence to classical thermodynamics, at least within the approximation of the second-order master equation \cite{ZBS}.

\smallskip {\em Stochastic limit.} If the probe can be described in terms of a position and a momentum variable (for example, a harmonic oscillator), then the reduced dynamics can be solved using path-integrals, through the Feynman-Vernon influence functional method \cite{IF}. Indeed, in this case our system is equivalent to a Caldeira-Leggett model \cite{CaLe} with  an ultra-ohmic  environment---see, Ref. \cite{HPZ}. In these systems, the influence functional technique allows one to pass to a semi-classical limit, of a particle moving under a stochastic force with correlator given by a component of the influence functional, known as the  {\em noise kernel}. Hence, one obtains the classical stochastic limit of a particle moving under a stochastic force.

In principle, this type of analysis is possible in the present system. We expect the noise kernel to depend explicitly on the function $N(\omega)$. Hence, the semiclassical limit of the probe would involve a Langevin equation for thermal noise that explicitly depends on the probe's velocity.

\section{Asymptotic states}

\subsection{Two level atom}
We consider the special case of a two-level atom of frequency $\Omega_0$. The Hamiltonian is $\hat{h} = \frac{1}{2} \Omega_0 \sigma_z$, and the coupling operator $\hat{A} = \hat{\sigma}_1$. Then, there are only two transition operators: $\hat{A}_{\Omega_0}$ corresponds to $\hat{\sigma}_-$ and   $\hat{A}_{-\Omega_0}$ corresponds to $\hat{\sigma}_+$. The master equation  takes the form
\begin{eqnarray}
\frac{\partial \hat{\rho}}{\partial \tau} = - i \Omega [ \hat{\sigma}_3, \hat{\rho}] + \Gamma_0  [N(\Omega_0)+1]  \left( \hat{\sigma}_-\hat{\rho}\hat{\sigma}_+  -\frac{1}{2}  \hat{\sigma}_+\hat{\sigma}_- \hat{\rho}  -  \frac{1}{2}  \hat{\rho}  \hat{\sigma}_+\hat{\sigma}_-  \right)\nonumber \\
+   \Gamma_0 N(\Omega_0)  \left(   \hat{\sigma}_+ \hat{\rho}\hat{\sigma}_-  -\frac{1}{2}  \hat{\sigma}_- \hat{\sigma}_+ \hat{\rho}  -  \frac{1}{2}  \hat{\rho} \hat{\sigma}_- \hat{\sigma}_+\right),
\end{eqnarray}
where $\Gamma_0 := \gamma(\Omega_0)$ is the decay coefficient for the atom in vacuum and $\Omega = \Omega_0 + 2 \Delta(\Omega_0)$ is the Lamb-shifted excitation frequency.

For a general initial pure state $|\psi_0\rangle= e^{i \phi} \cos\frac{\theta}{2}|1\rangle +\sin\frac{\theta}{2}|0\rangle$, the solution to the master equation is

 \begin{eqnarray}
\rho(\tau) = \frac{1}{2}
\bordermatrix{&              &              \cr
   & 1+e^{-\Gamma_0(2N+1)\tau}\cos\theta-\frac{1-e^{-\Gamma_0(2N+1)\tau}}{2N+1}
   & e^{-\frac{\Gamma_0}{2}(2N+1)\tau-i\Omega\tau + i \phi}\sin\theta \cr
                          & e^{-\frac{\Gamma_0}{2}(2N+1)\tau+i\Omega\tau - i \phi}\sin\theta
                          & 1-e^{-\Gamma_0(2N+1)\tau}\cos\theta +\frac{1-e^{-\Gamma_0(2N+1)\tau}}{2N+1} \cr}
\end{eqnarray}
where   $N = N(\Omega_0)$.

For an atom in the ground state ($\theta = \pi$), and for $\Gamma_0 \tau << 1$, $\hat{\rho}_{11}(t) = \Gamma_0 N \tau$, i.e., the excitation rate is equal to $\Gamma_0N(\Omega_0)$. This reproduces the result of Ref. \cite{CoMa} for the UdW coupling.

There is a unique asymptotic state
 \begin{eqnarray}
\rho(\tau) = \frac{1}{2 N(\Omega_0)+1 }
\bordermatrix{&              &              \cr
   & N(\Omega_0)
   & 0 \cr
                          & 0
                          & N(\Omega_0) + 1\cr}.
\end{eqnarray}
The expectation value of  energy is

\begin{eqnarray}
\langle \hat{h} \rangle = - \frac{ \Omega_0}{2 [2N(\Omega_0)+1]}.
\end{eqnarray}

The asymptotic state is not universal: it depends on $N(\Omega)$,  which depends on the type of coupling.


\subsection{Harmonic oscillator}
 For an harmonic oscillator of mass $m$ and frequency $\Omega_0$,  the Hamiltonian is $\hat{h} = \Omega_0 \hat{a}^{\dagger}\hat{a}$, and the coupling operator is $\hat{x} = \frac{1}{\sqrt{2m\Omega_0}}(\hat{a}+\hat{a}^{\dagger})$.  There are only two transition operators $\hat{A}_{\Omega_0} =  \frac{1}{\sqrt{2m\Omega_0}} \hat{a}$ and $\hat{A}_{-\Omega_0} =  \frac{1}{\sqrt{2m\Omega_0}}\hat{a}^{\dagger}$.

The master equation is

\begin{align*}
\frac{d}{d\tau}\hat{\rho}(\tau)
&=-i\Omega_0[\hat{a}^\dagger\hat{a},\hat{\rho}]+\Gamma_0\Big( N(\Omega_0) + 1 \Big)\Big(\hat{a}\hat{\rho}\hat{a}^\dagger-\frac{1}{2} \hat{a}^\dagger\hat{a}\hat{\rho}-\frac{1}{2}\hat{\rho}\hat{a}^\dagger\hat{a}\Big)+\\
& +\Gamma_0 N(\Omega_0) \Big(\hat{a}^\dagger\hat{\rho}\hat{a}-\frac{1}{2}\hat{a}\hat{a}^\dagger\hat{\rho} - \frac{1}{2}\hat{\rho}\hat{a}\hat{a}^\dagger\Big),
\end{align*}
where $\Gamma_0 = \frac{\gamma (\Omega_0)}{2m \omega_0}$.

There is a unique asymptotic state with matrix elements in the energy basis
\begin{eqnarray}
\rho_{nn'} =\frac{1}{N(\Omega_0)+1}\Bigg(\frac{N(\Omega_0)}{N(\Omega_0)+1}\Bigg)^n \delta_{nn'},
\end{eqnarray}
with mean energy
\begin{eqnarray}
\langle \hat{h} \rangle =  \Omega_0N(\Omega_0).
\end{eqnarray}

Again, the asymptotic state   depends only on $N(\Omega_0)$. It is unique for a given coupling, but it differs for different couplings.

 \subsection{Three Level Atom}
The last case considered here is that of a three-level atom with energy levels $|a\rangle, |b \rangle, |c \rangle$ and  associated energies $E_a < E_b < E_c$. For a dipole coupling with the EM field, one of the coupling constants for the three possible transitions must be zero. We consider the case that the transitions $a \leftrightarrow b$ is forbidden. We denote the transition $a \leftrightarrow c $ as $1$ with associated frequency $\Omega_1 = E_c -E_a $ and coupling constant $\lambda_1$, and the transition $b \leftrightarrow c $ as $2$ with associated frequency $\Omega_2 = E_c - E_b$ and coupling constant $\lambda_2$.

We choose the energy of the ground state $E_a = 0$, so that the self-Hamiltonian reads
\begin{eqnarray}
\hat{h} = (\Omega_1 - \Omega_2) |b\rangle \langle b| + \Omega_1 |c\rangle \langle c|.
\end{eqnarray}
The interaction term is
 \begin{equation}
\hat{V} =\Big(\lambda_1(\hat{s}_{1}+\hat{s}_{1}^{\dagger})+\lambda_2(\hat{s}_{2}+\hat{s}_{2}^{\dagger})\Big) \otimes \hat{O}(x)
 \end{equation}
where $\hat{s}_1 = |a\rangle \langle c|$ and $\hat{s}_2 = |b\rangle \langle c|$ are atomic transition operators. They satisfy $\hat{s}_1^2 = \hat{s}_2^2 = 0$.

There are 4 transition operators: $\hat{A}_{\Omega_1} = \lambda_1 \hat{s}_1, \hat{A}_{-\Omega_1} = \lambda_1 \hat{s}_1^{\dagger}, \hat{A}_{\Omega_2} = \lambda_2 \hat{s}_2$ and $\hat{A}_{-\Omega_2} = \lambda_2 \hat{s}_2^{\dagger}$. The master equation is

  \begin{align}
  \label{master3level}
  \begin{split}
  \frac{d}{d\tau}\hat{\rho}
  &=-i[\hat{h},\hat{\rho}]+\sum_{i=1}^{2}\Bigg( \Gamma_i (N_i+1)\Big(\hat{s}_{i}\hat{\rho}\hat{s}^{\dagger}_{i}-\frac{1}{2}\hat{s}^{\dagger}_{i}\hat{s}_{i}\hat{\rho} - \frac{1}{2}\hat{\rho}\hat{s}^{\dagger}_{i}\hat{s}_{i}\Big)+\\
  &+\Gamma_i  N_i \Big(\hat{s}^{\dagger}_{i}\hat{\rho}\hat{s}_{i}-\frac{1}{2}\hat{s}_{i}\hat{s}^{\dagger}_{i}\hat{\rho} -\frac{1}{2}\hat{\rho} \hat{s}_{i}\hat{s}^{\dagger}_{i}\Big)\Bigg),\\
  \end{split}
  \end{align}
where $\Gamma_i = \gamma(\Omega_i)$, $N_i = N(\Omega_i)$ and

\begin{eqnarray}
\hat{h} = (\Omega_1 +\Delta_1) |c\rangle \langle c| + (\Omega_1 - \Omega_2 + \Delta_2)|b\rangle \langle b|  - (\Delta_1 + \Delta_2) |a\rangle \langle a|
\end{eqnarray}
with $\Delta_i = \Delta (\Omega_i)$.

The stationary solution at late times is diagonal,

 \begin{align*}
 \rho_{aa}&=\frac{(N_1 + 1) N_2}{3 N_1 N_2 +N_1 +N_2}\\
 \rho_{bb}&=\frac{ N_1(N_2 + 1)}{3 N_1 N_2 +N_1 +N_2}\\
 \rho_{cc}&=\frac{ N_1 N_2}{3 N_1 N_2 +N_1 +N_2}.
 \end{align*}

As in the previous cases, the stationary solution does not depend on the strength of the interaction, but   it depends on the type of coupling through the parameters $N_i$.

\subsection{General systems}
In general, the existence of unique asymptotic solutions to the master equation (\ref{mastereq}) depends on the system Hamiltonian $\hat{h}$ and on the operators $\hat{A}_{\omega}$. The case of a non-degenerate Hamiltonian $\hat{h}$ is particularly important. Let us denote by $\epsilon_n$ the eigenvalues and by $|n\rangle$ the eigenvectors of $\hat{h}$, labeled by $n = 0, 1, 2, \ldots$ so that $\epsilon_n < \epsilon_{n'}$ for $n < n'$. Then, the diagonal elements of the density matric,
\begin{eqnarray}
p_n := \langle n|\hat{\rho}|n\rangle,
\end{eqnarray}
decouple from the off-diagonal ones, and they satisfy Pauli's master equation
\begin{eqnarray}
\frac{dp_n}{dt} = \sum_m (T_{nm}p_m - T_{mn}p_n), \label{pauli}
\end{eqnarray}
with transition rates
\begin{eqnarray}
T_{nm} := \gamma(|\epsilon_n - \epsilon_m|)|\langle m|\hat{A}|n\rangle|^2  \times \left\{ \begin{array}{cc} N(|\epsilon_n - \epsilon_m|) +1, & m > n\\ N(|\epsilon_n - \epsilon_m|) & n < m \end{array}  \right\}.
\end{eqnarray}
Asymptotic states correspond to probability vectors $p_m$ that are eigenvectors of $T_{nm}$.

The {\em detailed balance} condition is that in equilibrium, each independent summand in the right-hand-side of Eq. (\ref{pauli}) vanishes. It implies that,
\begin{eqnarray}
\frac{p_n}{p_m} = \frac{N(|\epsilon_n - \epsilon_m|)}{N(|\epsilon_n - \epsilon_m|) +1}, \label{ratio}
\end{eqnarray}
for $n > m$. Detailed balanced holds for all systems if $u = 0$, and it also holds for the systems studied in this section. We find it plausible that it holds for a generic non-degenerate Hamiltonian. A proof would require the application / generalization  of existing theorems about the asymptotic states of dynamical semigroups \cite{Spohn, Frigeiro}.

\subsection{Summary}
Our analysis of the master equation has revealed the following pattern.

\medskip

\noindent 1. There is a unique asymptotic state for each self-Hamiltonian $\hat{h}$.

\smallskip

\noindent 2. The asymptotic state depends only on the function $N(\omega)$. It does not depend on the relaxation time $\Gamma_0^{-1}$ of the system, i.e.,  on the strength of the system-reservoir coupling.

\smallskip

\noindent 3. The asymptotic state is not universal. It depends on the channel of interaction between the system and the thermal environment, i.e., on the composite operator $\hat{O}(x)$ that enters the coupling term (\ref{coupling}).

 \section{Thermodynamic characteristics}

\subsection{  The quantum probes as thermometers.}

An ensemble of  quantum systems interacting with a thermal reservoir is an elementary thermometer. The average energy of those systems in equilibrium serves as an empirical temperature for a bath. For example, an ensemble of harmonic oscillators of frequency $\Omega$ interacting with  thermal reservoir is characterized by an empirical temperature
\begin{eqnarray}
\theta(T, \Omega) = \frac{\Omega}{e^{\Omega/T} - 1}. \label{empiricalT}
\end{eqnarray}
The function (\ref{empiricalT}) satisfies the main criterion for an empirical temperature, namely,  it is an increasing function of the absolute temperature $T$.

The same reasoning applies to {\em moving} quantum systems in interaction  with a thermal reservoir. The average energy in the rest frame of the quantum system still serves as an empirical temperature that also depends on the rapidity  $u$. For a harmonic oscillator of frequency $\Omega$,
\begin{eqnarray}
\theta(T, \Omega, u ) = \Omega N(\Omega). \label{empiricalT2}
\end{eqnarray}
The function (\ref{empiricalT}) is also an increasing function of $T$, as can be seen by Eqs. (\ref{Nomega}) and (\ref{Nomega2}).

Hence, a physical system that can be used as a thermometer when at rest with respect to  a thermal reservoir remains a  thermometer when  moving. What changes is the explicit rule that connects the empirical temperature with the absolute temperature $T$ of the heat bath.

\subsection{Heat transfer}
The probe S can also be viewed as a thermodynamic system, subject to the   laws of thermodynamics. Consider a Markovian master equation of the form
\begin{eqnarray}
\frac{d}{dt} \hat{\rho} = - i [\hat{h}(t), \hat{\rho}] + {\cal L}[\rho], \label{mme}
\end{eqnarray}
where $\hat{h}(t)$ is the Hamiltonian of the probe and ${\cal L}$ is a super-operator of the Lindblad-Kossakowski (LK) type that generates non-unitary evolution. Eq. (\ref{mme}) leads to a non-equilibrium formulation of the first law of thermodynamics \cite{Alicki}
\begin{eqnarray}
\frac{d}{dt}E =  q - P,
\end{eqnarray}
where $E = \langle \hat{h}\rangle$ is the internal energy, $P := - Tr(\hat{\rho}\frac{d\hat{h}}{dt})$ is the power provided to the system, and
\begin{eqnarray}
q= Tr ( {\cal L}[\hat{\rho}]\hat{h})
\end{eqnarray}
is the total heat current.

In the present context, $\frac{d\hat{h}}{dt} = 0$, hence, $P = 0$. All changes in the internal energy are due to the heat current. For the master equation (\ref{mastereq}), the heat current is
\begin{eqnarray}
q =   \sum_{\omega > 0} \gamma(\omega) \omega\left\{ N(\omega)\langle \hat{A}_{\omega} \hat{A}_{\omega}^{\dagger}\rangle - [N(\omega)+1] \langle \hat{A}^{\dagger}_{\omega} \hat{A}_{\omega}\rangle \right\} \label{heatcur}
\end{eqnarray}
In equilibrium, $q = 0$. Hence, the equilibrium state satisfies  $N(\omega)\langle \hat{A}_{\omega} \hat{A}_{\omega}^{\dagger}\rangle = [N(\omega)+1] \langle \hat{A}^{\dagger}_{\omega} \hat{A}_{\omega}\rangle$ for all $\omega$.


Let the initial state of the moving system be thermal at temperature   $T_0$ in  the CoM frame.  We proceed to evaluate  the total heat transferred from the reservoir to the system,
\begin{eqnarray}
\Delta Q = E(\infty) - E(0),
\end{eqnarray}
 and the heat current (\ref{heatcur}) as a function of time.

\medskip

 For a qubit,
\begin{eqnarray}
\Delta Q = \frac{\Omega_0}{(2n_0+1)[2N(\Omega_0)+1]} [ N(\Omega_0)-n_0],
\end{eqnarray}
where $n_0 = (e^{\frac{\Omega_0}{T_0}}-1)^{-1}$, and
\begin{eqnarray}
q = \frac{1}{2n_0+1}\Gamma_0 \Omega_0 e^{-\Gamma_0[1+2N(\Omega_0)]\tau}[N(\Omega_0)-n_0].
\end{eqnarray}

\medskip

For a harmonic oscillator,
\begin{eqnarray}
\Delta Q =  \Omega_0 [ N(\Omega_0)-n_0],
\end{eqnarray}
and
\begin{eqnarray}
q =  \Gamma_0 \Omega_0 e^{-\Gamma_0 \tau}[ N(\Omega_0)-n_0].
\end{eqnarray}

The systems above have a consistent thermodynamical behavior\footnote{The reader may worry that a single harmonic oscillator or a single qubit is not a thermodynamic system. However, the same results hold for a collection of $N$ qubits or harmonic oscillators, with mutual  interactions much weaker that the interaction with the bath, i.e., for a `dilute gas' of qubits or harmonic oscillators, which is a thermodynamic system for $N >> 1$.}.
 If $\Delta Q > 0$, then the initial state is colder than the final, there is positive heat transfer, and the heat current  is positive at all times. An analogous statement holds for $\Delta Q < 0$.

Heat flows even when $T = T_0$. In Fig. 2, we plot the heat transfer $\Delta Q$ in the qubit system for $T = T_0$, as a function of $u$. We note that $\Delta Q < 0$  for $\beta \omega < 1$. From the perspective of the rest frame of $S$, a moving heat bath at sufficiently high temperature is always colder than a stationary one.

\begin{figure}[H]
    \centering
   {{\includegraphics[width=11cm]{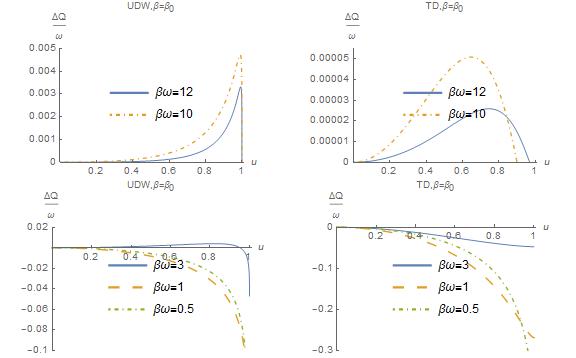} }}%
  \caption{  \small  The heat transfer $\Delta Q$ for a qubit at temperature $T_0 = T$ as a function of $u$ for different values of $\beta \omega$. The left plot describes a qubit interacting with the field bath through the UdW coupling and the right one  describes a qubit interacting with the field bath through the TD coupling.}
\end{figure}

In some axiomatic approaches to thermodynamics \cite{Serrin}, the notion of `hotness' is introduced as a primitive structure. Hotness is an order relation $\preceq$ on the set $\Gamma$ of thermodynamic states:  $A \preceq B$ if there is (non-negative) heat flow from $A$ to $B$ when two bodies on states $A$ and $B$ are brought in contact.   The order relation is total, i.e., for any pair of states, either $A \preceq B$ or $B \preceq A$. This property enables us to express $\preceq$ in terms of the inequality relation of real numbers, and hence,   to introduce the notion of temperature.

The results of this section strongly suggest that the notion of hotness could also be meaningful on an extended thermodynamic state space $\tilde{\Gamma}$ that  includes information about the motion of the system's CoM. This means that the order relation  $\preceq$ can be extended to $\tilde{\Gamma}$.

Furthermore, as the sign of heat flow depends only on the sign of the difference $ N(\Omega_0)-n_0$, our results are compatible with the idea that $\preceq$ is a total order also on $\tilde{\Gamma}$. However, this assertion is too strong: it implies a universal notion of temperature that also incorporates the effects of the CoM motion. We have not found such a candidate for temperature in our study. Further research is necessary in order to understand the properties of the proposed notion of hotness.

\subsection{Zero-th law of thermodynamics.}
The existence of a unique asymptotic state for each Hamiltonian $\hat{h}$ is a necessary consequence of the zero-th law of thermodynamics. Any system in contact with a thermal reservoir of temperature $T$ at rest reaches equilibrium at temperature $T$.

Two states A and B, are in  thermal equilibrium ($A \sim B$), if $A \preceq B$ and $B \preceq A$, i,e., if there is no heat flow when they are brought into contact.

Consider two systems in states A and B,  in thermal equilibrium with the same reservoir C. If the systems are removed  from the reservoir and they are brought into thermal contact with each other, there is no heat flow. This manifests  the crucial feature of the zero-th law of thermodynamics, namely: If $A \sim  C$ and $B \sim C$, then $A \sim  B$. This {\em transitivity property} is inherent in the definition of the order relation $\preceq$.

Next, we consider two  systems in states A and B that move with the same velocity $v$ with respect to a thermal reservoir (system C). System A interacts with the reservoir through the UdW coupling, while  system B interacts  through the TD coupling.    A and B reach the equilibrium state, and then they are removed from the reservoir. They are at rest with each other and they are brought into contact.

Systems A and B have different energy, while their self-Hamiltonians are identical, and the values of other extensive quantities can be taken to be equal. In usual thermodynamics, the value of energy and the other extensive quantities uniquely define all thermodynamic potentials, and hence, the notion of  thermal equilibrium. This would imply that the systems A and B are not in thermal equilibrium with each other:
 they will exchange heat until they are brought into a new equilibrium state.   There is no transitivity for three systems A, B, and C, if one of them is in motion with respect to the other.

Hence, in order to compare systems that move with respect to each other, we must either abandon the zero-th law of thermodynamics, or  introduce additional variables to describe thermodynamic states.  The zero-th law of thermodynamics is a consequence of the partial ordering relation $\preceq$ on $\tilde{\Gamma}$, which is reasonably well justified by our previous analysis. For this reason, we believe that the zero-th law is preserved and that the  thermodynamic description of moving systems requires the introduction of  additional variables, beyond the components of the four velocity that are employed in the van-Kampen-Israel formalism \cite{vankampen, Israel}. Indeed, such variables emerge from our account of the second law of thermodynamics.


\subsection{Directional averaging and directional temperature}
Next, we present an important identity that is satisfied by the function $N(\omega)$ and it is crucial to the thermodynamic interpretation of the master equation (\ref{mastereq}). To this end, we first define the notion of directional averaging. Let $V$ be the set of null vectors $p_{\mu} = (|{\pmb p}|, {\pmb p})$, and  $f: {\pmb R} \rightarrow {\pmb R}$ a function of $\omega = |{\pmb p}|$. Consider a Lorentz boost of rapidity $u$ along a direction ${\pmb n}$. In the boosted frame, the energy is $\omega' = \omega \cosh u - {\pmb p}\cdot {\pmb n} \sinh u = \omega (\cosh u - {\pmb s} \cdot {\pmb n} \sinh u)$, where ${\pmb s} = {\pmb p}/\omega$. We define the directional averaging $\langle f(\omega)\rangle_u$ as
\begin{eqnarray}
\langle f(\omega)\rangle_u  := \int d \mu({\pmb s}) f(\omega'), \label{dirav}
\end{eqnarray}
where $d \mu({\pmb s})$ is the invariant, normalized measure on the unit sphere. The directional averaging of $f$ does not depend on the direction ${\pmb n}$ of the boost. Defining $\xi:= {\pmb s} \cdot {\pmb n}$, Eq. (\ref{dirav}) becomes
\begin{eqnarray}
\langle f(\omega)\rangle_u = \frac{1}{2} \int_{-1}^1 d \xi  f[\omega(\cosh u - \xi \sinh u)], \label{dirav2}
\end{eqnarray}

With this definition,
\begin{eqnarray}
N_{UdW}(\omega) = \langle n(\omega)\rangle_u,
\end{eqnarray}
where $n(\omega) = (e^{\beta \omega}-1)^{-1}$ is the Planck distribution. Indeed,
\begin{eqnarray}
\langle n(\omega)\rangle_u = \frac{1}{2}  \sum_{m=1}^{\infty} \int_{-1}^1 d \xi e^{-m\omega(\cosh u - \xi \sinh u)} \nonumber \\ = \frac{1}{2\beta \omega \sinh u} \sum_{m=1}^{\infty}\Big(\frac{e^{-m\beta \omega e^{-u}}}{m}-\frac{e^{-m \beta \omega e^{u}}}{m}\Big) = N_{UdW}(\omega).
\end{eqnarray}

Similarly, we find that
\begin{eqnarray}
N_{TD}(\omega) = \frac{3\langle  \omega^2 n(\omega)\rangle_u}{\omega^2[1+2\cosh(2u)]}.
\end{eqnarray}

We define the directional temperature
\begin{eqnarray}
T_{\xi}:=\frac{T}{\cosh u - \xi \sinh u},
\end{eqnarray}
or equivalently $\beta_\xi:= \beta (\cosh u - \xi \sinh u)$. For each $u$, $Te^{-|u|} \leq T_{\xi} \leq Te^{-|u|} $.

Then, $N(\omega)$ can be written as a weighted average of the Planck distribution for varying temperatures $T_{\xi}$,
\begin{eqnarray}
N(\omega) = \frac{1}{2} \int_{-1}^1 d\xi \frac{w(\xi)}{e^{\beta_\xi \omega}-1},  \label{Nweight}
\end{eqnarray}
where  $w(\xi)$ is a probability density on $[-1, 1)$ that depends on the coupling:
\begin{eqnarray}
w_{UdW}(\xi) &=& 1 \\
w_{TD}(\xi) &=&  \frac{3(\cosh u - \xi \sinh u)^2}{1+2 \cosh(2u)}.
\end{eqnarray}

\subsection{Second law of thermodynamics}
Eq. (\ref{Nweight}) implies that the master equation (\ref{mastereq}) can be expressed as
\begin{eqnarray}
\frac{\partial \hat{\rho}}{\partial \tau} = - i [ \hat{h} + \hat{h}_{LS}, \hat{\rho}] + \frac{1}{2} \int_{-1}^1 d\xi w(\xi) {\cal L}_{\beta_\xi}[\hat{\rho}], \label{mastereq2}
\end{eqnarray}
where
\begin{eqnarray}
 {\cal L}_{\beta_{\xi}}[\hat{\rho}] =  \sum_{\omega > 0 }  \frac{\gamma(\omega)}{1- e^{-\beta_{\xi} \omega}}  \left[ \hat{A}_{\omega}\hat{\rho}\hat{A}^{\dagger}_{\omega} -\frac{1}{2}  \hat{A}^{\dagger}_{\omega}\hat{A}_{\omega}\hat{\rho}  -  \frac{1}{2}  \hat{\rho} \hat{A}^{\dagger}_{\omega} \hat{A}_{\omega}\right]\nonumber \\
+     \sum_{\omega > 0 }  \frac{\gamma(\omega)}{e^{\beta_{\xi} \omega} - 1}  \left[ \hat{A}^{\dagger}_{\omega}\hat{\rho}\hat{A}_{\omega} -\frac{1}{2}  \hat{A}_{\omega}\hat{A}^{\dagger}_{\omega}\hat{\rho}  -  \frac{1}{2}  \hat{\rho} \hat{A}_{\omega} \hat{A}^{\dagger}_{\omega}\right],
\end{eqnarray}
is a LK  map for a thermal reservoir at inverse temperature $\beta_{\xi}$.

Hence, Eq. (\ref{mastereq}) can be interpreted as a master equation for  a system {\em in contact with a continuum of different reservoirs at temperatures $T_{\xi}$}.

For a thermal reservoir at temperature $T$, the thermodynamic entropy $S$ coincides with the von Neumann entropy $S_{vN} := - Tr \hat{\rho} \log{\rho}$, and it satisfies the balance equation
\begin{eqnarray}
\frac{dS}{dt} - \beta q_{\beta} = \sigma_{\beta},
\end{eqnarray}
where  $q_{\beta} = ({\cal L}_{\beta}[\hat{\rho}] \hat{h})$ is the heat flux, and $\sigma$ is the total entropy production \cite{SL78}. The LK operator ${\cal L}_{\beta}$ for a thermal reservoir must have a Gibbsian equilibrium state $\hat{\rho}_{\beta} = \frac{e^{-\beta \hat{h}}}{Tr e^{-\beta \hat{h}}}$. Then, entropy production is given by \cite{SL78}
\begin{eqnarray}
\sigma_{\beta} = - Tr\left({\cal L}_{\beta}[\hat{\rho}](\log \hat{\rho} - \log \hat{\rho}_{\beta}\right) \geq 0. \label{sigmab}
\end{eqnarray}
Note that $\sigma_{\beta} = - \beta \frac{dF}{dt}$, where $F$ is the Helmholz free energy of the total system that includes the probe and the  reservoir at temperature $\beta^{-1}$.

In Eq. (\ref{mastereq2}), the LK map is a weighted  average of the LK maps for different thermal reservoirs. Since entropy production is a linear functional of the LK map, we can express the entropy production associated to  Eq. (\ref{mastereq2}) as an average of $\sigma_{\beta}$, Eq. (\ref{sigmab}),
\begin{eqnarray}
\sigma = \frac{1}{2} \int_{-1}^1 d\xi w(\xi) \sigma_{\beta_{\xi}}.
\end{eqnarray}
It follows that
\begin{eqnarray}
\sigma =  - Tr\left({\cal L}[\hat{\rho}] \log \hat{\rho}\right) - \frac{1}{2} \int_{-1}^1 d\xi w(\xi) \beta_{\xi} Tr\left({\cal L}_{\beta_{\xi}}[\hat{\rho}] \hat{h}\right).
\end{eqnarray}

The total produced entropy is given by
\begin{eqnarray}
\Delta S_{tot} = S_{vN}[\hat{\rho}(\infty)] - S_{vN}[\hat{\rho}(0)] -  \frac{1}{2} \int_{-1}^1 d\xi w(\xi) \beta_{\xi} [E_{\xi} - E(0)],
\end{eqnarray}
where $E_{\xi} = Tr (\hat{\rho}_{\beta_{\xi}}\hat{h})$.

Plots of the entropy production as a function of $t$ are given in Fig.3, and of the total produced entropy as a function of $u$ in Fig. 4.
\begin{figure}[H]
    \centering
   {{\includegraphics[width=11cm]{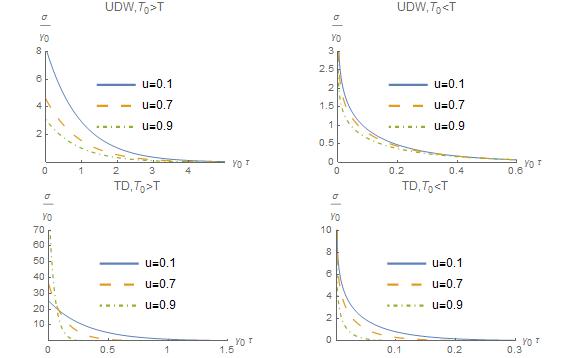}}}%
 \caption{\small The dimensionless entropy production $\sigma /\gamma_0$ of a qubit as a function of dimensionless time $\gamma_0 t$. For the UdW coupling, $\gamma_0 = \frac{\lambda^2}{2\pi} \omega$, while for the TD coupling $\gamma_0 = \frac{\lambda^2}{6\pi} \omega^3$. The temperatures $T_0$ and $T$ alternate  values $33 \omega^{-1}$ and $ \omega^{-1}$.}
\end{figure}
\begin{figure}[H]
\label{N}%
    \centering
   {{\includegraphics[width=11cm]{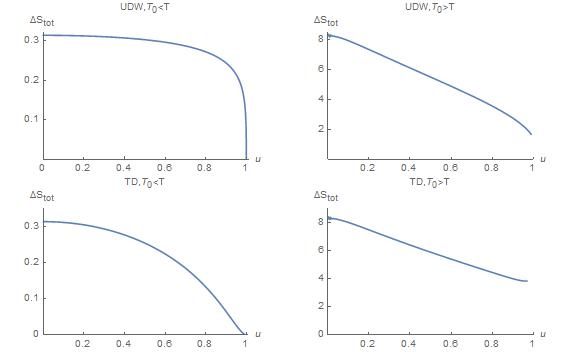}}}%
 \caption{\small Total produced entropy  $\Delta S_{tot}$   as a function of the rapidity $u$. The temperatures $T_0$ and $T$ alternate  values $33 \omega^{-1}$ and $ \omega^{-1}$.}
\end{figure}

 \section{Relativistic transformations for thermodynamic states}

\subsection{Relativistic transformation of Gibbs states}
Eq. (\ref{mastereq2}) implies the following equivalence for an quantum open system. The non-equilibrium dynamics of a heat bath at temperature $T_0$ moving with velocity $-v = - \tanh u$ is equivalent to the non-equilibrium dynamics of a continuum of heat baths at rest,  and with temperatures $T_{\xi} \in [T_0e^{-|u|}, T_0e^{|u|}]$. The different baths are weighted by a system-specific probability distribution $w(\xi)$.

This  leads to the conjecture that the effect of a Lorentz boost with rapidity $u$ on a Gibbs state $\hat{\rho}_{\beta}$ is a convex combination of Gibbs states $\hat{\rho}_{\beta_\xi}$ with a weight $w(\xi)$,
\begin{eqnarray}
\hat{\rho}_{\beta} \rightarrow \frac{1}{2} \int_{-1}^1 d \xi w(\xi) \hat{\rho}_{\beta_\xi}. \label{conjj}
\end{eqnarray}

 Obviously, this transformation is different from the unitary  transformation of the Gibbs state through a unitary representation of the Lorentz group $\hat{U}(\Lambda)$,
\begin{eqnarray}
\hat{\rho}_{\beta} \rightarrow \hat{U}^{\dagger}(\Lambda)\hat{\rho}_{\beta}\hat{U}(\Lambda).
\end{eqnarray}
 The latter transformation does not change the thermodynamic description of the system, as the partition function is invariant under unitary transformations.

   Eq. (\ref{conjj}) should not be taken literally. It refers only to the thermodynamic level of description, i.e., to properties of the Gibbs state that relevant to  thermodynamic properties.  A more precise version of (\ref{conjj}) is the following. Let $B_{th}$ be the subset of operators on the Hilbert space that describe thermodynamic  observables. Then, the
   that the expectation value $\langle \hat{A}\rangle_u$ for any $\hat{A} \in B_{th} $ in the moving frame should be given by
\begin{eqnarray}
\langle \hat{A}\rangle_u = \frac{1}{2} \int_{-1}^1 d \xi w(\xi) Tr (\hat{\rho}_{\beta_\xi} \hat{A}). \label{conjj2}
\end{eqnarray}
Observables outside $B_{th}$ are not constrained\footnote{Note that in many approaches to non-equilibrium thermodynamics, the microcanonical state is obtained as an asymptotic state, only with respect to a small subset of macroscopic observables that have a thermodynamic interpretation---see, for example \cite{OP}.}.

 Eq. (\ref{conjj2}) is also supported by the following properties of the field correlation functions. Consider the thermal Hadamard function of the massless scalar field, defined by $G_{\beta}(x ) = \frac{1}{2} Tr (\hat{\rho}_{\beta} \{\hat{\phi}(x),\hat{\phi}(0)\})$,
\begin{eqnarray}
G_{\beta}(t, {\pmb r}) =- \frac{1}{4\pi^2} \sum_{n=-\infty}^{\infty} \frac{1}{(t + i n \beta)^2 - {\pmb r}^2} \label{Hadamar}
\end{eqnarray}
Eq. (\ref{Hadamar}) satisfies the Kubo-Martin-Schwinger (KMS) condition
\begin{eqnarray}
G_{\beta}(t - i \beta, {\pmb r}) = G_{\beta}(t , {\pmb r}). \label{KMS}
\end{eqnarray}
Next, consider the Lorentz transformed Hadamard function $G^{u}_{\beta}(x) := G_{\beta}(\Lambda_u^{-1}x)$, where $\Lambda_u$ is a Lorentz boost of  rapidity $u$ in the direction 1,
\begin{eqnarray}
\Lambda_u(t,x_1, x_2, x_2) = (t \cosh u - x_1 \sinh u, x_1 \cosh u - t \sinh u, x_2, x_3).
\end{eqnarray}
We obtain
\begin{eqnarray}
G^u_{\beta}(t, {\pmb r}) =- \frac{1}{4\pi^2} \sum_{n=-\infty}^{\infty} \frac{1}{t^2 - {\pmb r}^2 - n^2 \beta^2  +2in\beta (t \cosh u +x_1 \sinh u)} \label{Hadamar2}
\end{eqnarray}
Eq. (\ref{Hadamar2}) does not satisfy the KMS condition. However, the probabilities for local measurements of the field are evaluated for ${\pmb r} = 0$,
\begin{eqnarray}
G^u_{\beta}(t, 0) =- \frac{1}{4\pi^2} \sum_{n=-\infty}^{\infty} \frac{1}{t^2  - n^2 \beta^2  +2in\beta t \cosh u  }. \label{Hadamar3}
\end{eqnarray}
Then, it is straightforward to prove that
\begin{eqnarray}
G^u_{\beta}(t, 0) = \frac{1}{2} \int_{-1}^{1} d \xi G_{\beta_{\xi}}(t, 0), \label{avvv}
\end{eqnarray}
i.e., the boosted Hadamard function is {\em a convex combination of functions that satisfy the KMS condition}. The same holds for the Hadamard function for the TD coupling.

Eq. (\ref{avvv}) justifies Eq. (\ref{conjj2}) for observables of the form
\begin{eqnarray}
\hat{A}_{\pmb x} = \int dt dt' a(t,t') \hat{\phi}(t, {\pmb x}) \hat{\phi}(t', {\pmb x}).
\end{eqnarray}
This class of observables includes ones that correspond to localized measurements of particle number and energy \cite{AnSav11, AnSav12, AnSav19, AnSav20}.  These observables can be used to define thermodynamic variables like particle number density or energy density.

\subsection{An extended thermodynamic space}
In Sec. 4.3, we showed that the applicability of the zero-th law of thermodynamics to moving systems requires a significant extension of the thermodynamic state space. Here, we present such an extension that is consistent with the relativistic transformation law discussed in Sec. 5.1.

First, we recall how the thermodynamic state space and the thermodynamic potentials of a quantum system are constructed from the canonical distribution. Let $\hat{H}(X)$ be the Hamiltonian that of the system in the CoM frame; $X$ are thermodynamic control parameters like volume, or external fields. The thermodynamic state space $\Gamma$ in the Helmholz representation has elements $(\beta, X)$, and it is in one-to-one correspondence with the set of all Gibbs states
$\hat{\rho}(\beta, X) = \frac{e^{-\beta \hat{H}(X)}}{Z(\beta, X)}$, where $Z(\beta, X) = e^{-\beta \hat{H}(X)}$ is the partition function. The thermodynamical potentials on $\Gamma$ can be derived by identifying  the expectation  $\langle \hat{H}\rangle$ with the internal energy, and  the von Neumann entropy $S_{vN} = - Tr \hat{\rho}\ln \hat{\rho}$ with the thermodynamic entropy.

Next, we consider the space $\tilde{\Gamma}$ that is a convex hull of $\Gamma$,  constructed through the Gibbs states. That is, $\tilde{\Gamma}$ is the set of all density matrices $\hat{\rho}= \sum_i c_i \hat{\rho}(\beta_i, X_i)$, for all sequences $\{c_i\}$ such that $0 \leq c_i \leq 1$ and $\sum_i c_i = 1$. Again, we construct the thermodynamic potentials on $\tilde{\Gamma}$ by identifying $\langle \hat{H}\rangle$ with the internal energy  and $S_{vN}$ with the thermodynamic entropy.

By construction,  the map (\ref{conjj}) is well-defined on $\tilde{\Gamma}$. It maps all extreme points of $\tilde{\Gamma}$ (i.e., Gibbs states), to points in the interior of $\tilde{\Gamma}$. Its action can be extended to any $\hat{\rho} \in \tilde{\Gamma}$ by linearity.

The first law of thermodynamics on $\tilde{\Gamma}$ is well-defined, since internal energy and entropy are well-defined. Our construction of $\tilde{\Gamma}$  is compatible with the analysis of the zero-th law in Sec. 4.3, and it enables a representation of the Lorentz boosts as discussed in Sec. 5.1. It is therefore a natural candidate for an extended thermodynamic state space that also takes into account the CoM motion of thermodynamic systems.

It is possible that the extended state space $\tilde{\Gamma}$ constructed here is larger than needed. For example, the map (\ref{conjj}) is well defined on a subset of $\tilde{\Gamma}$ that consists of states of the form $\hat{\rho}= \sum_i c_i \hat{\rho}(\beta_i, X)$, i.e., the convex combinations involve only different values of temperature and not of $X$. However, our preliminary analysis serves to highlight the key point of the analysis of the 2nd law in Sec. 4.E:  thermodynamic transformations between different Lorentz frames can be implemented in terms of convex combinations of Gibbs states.

\section{Conclusions}
We analysed the quantum thermodynamics of moving systems in interaction with a heat bath. We showed that these systems are well-behaved thermodynamically, in the sense that they have a consistent notion of heat flow. There is no relativistic rule for transformation of temperature, however, a moving heat bath is equivalent to a continuum of stationary heat baths, as far as the non-equilibrium dynamics of the system is concerned. This led us to the proposal of an extended thermodynamic state space in which the Lorentz transformations can be well implemented.

Our results are derived using specific models, rather than general mathematical principles. It is therefore necessary to develop models that deal with more elaborate cases. We must consider other  types of thermal bath, for example, relativistic gases of massive particles. Furthermore, we must generalize the present results to extended quantum systems that are not defined by a point-like trajectory. This is essential for incorporating observables like volume and pressure in the thermodynamic description. This will enable us to connect directly with the traditional accounts of Lorentz transformation for thermodynamic variables.
One possible approach towards this goal is to adopt the methods and techniques used in Ref. \cite{AnSav19} for non-pointlike detectors.

If our conjecture about the relativistic transformation rule is confirmed by other models, it will be necessary to look for a more fundamental justification. This could be provided by  an  analysis of QFT two-point functions like Eq. (\ref{Wightman0}) for general composite operators $\hat{O}(x)$ and KMS states. Furthermore, it is important to consider possible experimental implementations of the models presented here.

\end{document}